\documentclass[english,journal,12pt,onecolumn,draftclsnofoot]{IEEEtran}
\usepackage[latin9]{inputenc}
\usepackage{color}
\usepackage{array}
\usepackage{booktabs}
\usepackage{textcomp}
\usepackage{multirow}
\usepackage{amsmath}
\usepackage{amssymb}
\usepackage{graphicx}

\makeatletter

\providecommand{\tabularnewline}{\\}
\newcommand{\lyxdot}{.}

\usepackage[center]{caption}
\usepackage[font=footnotesize]{caption}
\@ifundefined{showcaptionsetup}{}{%
 \PassOptionsToPackage{caption=false}{subfig}}
\usepackage{subfig}
\makeatother

\usepackage{babel}
\begin{document}
\title{\vspace{-4mm}
Low Complexity Neural Network Structures for Self-Interference Cancellation
in Full-Duplex Radio}
\author{Mohamed Elsayed, \textit{Student Member, IEEE}, Ahmad A. Aziz El-Banna,
\textit{Member, IEEE}, Octavia A. Dobre, \textit{Fellow, IEEE}, Wanyi
Shiu, and Peiwei Wang\vspace{-9.5mm}
\thanks{This work is supported by Huawei Canada Research Centre, Huawei Technologies,
Canada.} \thanks{M. Elsayed, A. A. A. El-Banna, and O. A. Dobre are with the Faculty
of Engineering and Applied Science, Memorial University, St. John\textquoteright s,
NL A1B 3X5, Canada (e-mail: \{memselim, aaelbanna, odobre\}@mun.ca).} \thanks{A. A. A. El-Banna is also on leave from the Faculty of Engineering
at Shoubra, Benha University, Banha, Egypt.} \thanks{W. Shiu and P. Wang are with Huawei Canada Research Centre, Huawei
Technologies Canada Co., Ltd., Ottawa, ON K2K 3J1, Canada (e-mail:
\{wanyi.shiu, peiwei.wang\}@huawei.com).}}
\maketitle
\begin{abstract}
Self-interference (SI) is considered as a main challenge in full-duplex
(FD) systems. Therefore, efficient SI cancelers are required for the
influential deployment of FD systems in beyond fifth-generation wireless
networks. Existing methods for SI cancellation have mostly considered
the polynomial representation of the SI signal at the receiver. These
methods are shown to operate well in practice while requiring high
computational complexity. Alternatively, neural networks (NNs) are
envisioned as promising candidates for modeling the SI signal with
reduced computational complexity. Consequently, in this paper, two
novel low complexity NN structures, referred to as the ladder-wise
grid structure (LWGS) and moving-window grid structure (MWGS), are
proposed. The core idea of these two structures is to mimic the non-linearity
and memory effect introduced to the SI signal in order to achieve
proper SI cancellation while exhibiting low computational complexity.
The simulation results reveal that the LWGS and MWGS NN-based cancelers
attain the same cancellation performance of the polynomial-based canceler
while providing 49.87\% and 34.19\% complexity reduction, respectively. 
\end{abstract}

\vspace{-2mm}


\vspace{-7mm}

\section{Introduction }

\def\figurename{Fig.}
\def\tablename{TABLE}

The recent advancements in wireless technology impose a tremendous
increase in the number of devices that are required to satisfy the
ascending demand for high data rates communication. This high increase
leads to an undeniable fact that some levels of saturation in the
available frequency resources will be reached. Therefore, using efficient
methods for sharing the spectrum resources is eagerly mandated for
the next generations of wireless systems, such as beyond the fifth-generation
{[}\ref{Toward 6G networks}{]}. The full-duplex (FD) technology has
emerged as a promising remedy for the spectrum congestion problem
by providing an efficient way for spectrum sharing. In the FD systems,
the data is transmitted and received at the same time slot and in
the same band of frequency {[}\ref{K.-E.-Kolodziej,}{]}. Sharing
the spectrum resources simultaneously has the potential of doubling
the spectral efficiency of the FD systems. However, this in turn,
gives rise to a substantial problem known as the self-interference
(SI), which occurs when the transmitter\textquoteright s signal is
leaked into the FD receiver. As such, canceling the SI signal at the
receiver is deemed the main challenge against the practical deployment
of FD systems {[}\ref{A.-Sabharwal,-P.}{]}, {[}\ref{Le-Ngoc}{]}. 

For typical FD systems, the SI signal could be 110 dB larger than
the desired signal of interest at the receiver {[}\ref{G. Agrawal, S. Aniruddhan, and R. K. Ganti}{]}.
Therefore, if not efficiently eliminated, the SI signal may saturate
the receiver's analog components, such as the analog-to-digital converter
(ADC) and the low-noise amplifier (LNA) {[}\ref{K.-E.-Kolodziej,}{]}.
Existing methods for SI cancellation employ analog domain cancellation
techniques, which are performed either passively using the physical
separation between the transmit and receive antennas or actively by
injecting a cancellation waveform into the propagation path of the
received signal {[}\ref{A.-Sabharwal,-P.}{]}. However, the analog
cancellation techniques are not usually able to completely eliminate
the SI signal at the receiver. Hence, the residual amount of the SI
signal is further suppressed with the help of digital domain cancellation
{[}\ref{Z. Zhang, K. Long, A. V. Vasilakos, and L. Hanzo}{]}. For
that, the digital transmitted signal is subtracted from the received
signal to perform the SI cancellation. The digital cancellation procedure
seems to be an easy task in theory; however, it is hard to be realized
in practice due to the non-linear distortion caused by the various
parts of the transceiver, such as the power amplifier (PA), IQ mixer,
ADC, and digital-to-analog converter (DAC) {[}\ref{D.-Korpi,-L.}{]}.
This distortion makes the SI signal entirely different from the digital
transmitted signal and raises a challenge for the perfect elimination
of the SI signal at the receiver. Typically, the polynomial model
is used for modeling the non-linearities caused by different parts
of the FD transceiver. The polynomial model works properly in practice
while suffering from high computational complexity {[}\ref{A.-Balatsoukas-Stimming,-"Non-li}{]}. 

Recently, neural networks (NNs) have received remarkable research
interest from communication community experts due to their advantages
in modeling the non-linearities with reduced computational complexity
{[}\ref{A.-Balatsoukas-Stimming,-"Non-li}{]}--{[}\ref{A.-T.-Kristensen,}{]}.
In {[}\ref{A.-Balatsoukas-Stimming,-"Non-li}{]}, the authors introduce
a real-valued feed-forward NN (RV-FFNN) to model the SI signal. Further,
the hardware implementation of this NN-based canceler is presented
in {[}\ref{Y. Kurzo, A. T. Kristensen, A. Burg, A. Balatsoukas-Stimming}{]}.
The same research group proposes the complex-valued FFNN (CV-FFNN)
to perform the SI cancellation, and shows that the CV-FFNN could achieve
the same cancellation as the RV-FFNN with a reduced number of floating-point
operations (FLOPs) {[}\ref{A.-T.-Kristensen,}{]}. In addition, in
{[}\ref{A.-T.-Kristensen,}{]}, the recurrent NN (RNN) is introduced
for SI cancellation due to its capability to model data sequences;
it has been shown that the RNN is not a proper candidate solution
for the SI cancellation problem due to its high computational complexity.
To the best of our knowledge, the research works of {[}\ref{A.-Balatsoukas-Stimming,-"Non-li}{]}
and {[}\ref{A.-T.-Kristensen,}{]} represent the few attempts that
target the application of NNs for SI cancellation in FD systems, and
there is a scarcity of contributions in this field. 

Subsequently, in this paper, two novel NN structures, referred to
as the ladder-wise grid structure (LWGS) and moving-window grid structure
(MWGS), are proposed. The aim of these structures is to model the
SI signal with low computational complexity. The proposed NNs exploit
a grid topology in which only partial connections among the different
neurons in the input and hidden layers are utilized to model the SI
signal with reduced computational complexity. In addition, the proposed
methods aim to learn the memory effect introduced to the SI signal
in order to efficiently perform the SI cancellation. The numerical
simulations substantiate the validity of the proposed LWGS and MWGS
NN-based cancelers as they achieve the same cancellation performance
of the polynomial canceler while providing 49.87\% and 34.19\% reduction
in the number of FLOPs, respectively. Besides, the proposed LWGS and
MWGS  outperform the state-of-the-art NN-based cancelers in terms
of computational complexity. 

\begin{figure}
\begin{centering}
\includegraphics[scale=0.26]{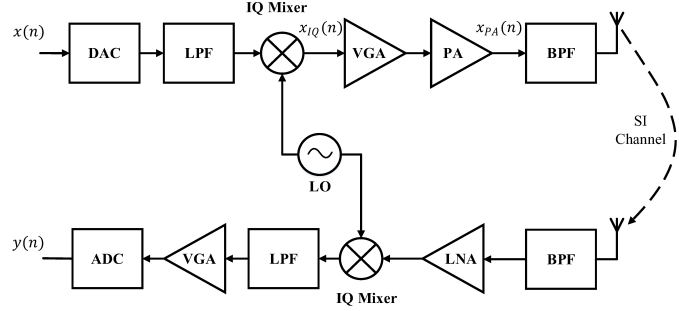}
\par\end{centering}
\centering{}\caption{\label{fig:Full-duplex-transceiver-system}Full-duplex transceiver
system model. }
\end{figure}

\vspace{-3mm}

\section{Full-Duplex System Model }

The system model of the FD transceiver is depicted in Fig. \ref{fig:Full-duplex-transceiver-system}.
In this paper, the polynomial model is used to approximate the SI
signal. Therefore, we follow the stipulated assumption in {[}\ref{D.-Korpi,-L.}{]}
that the IQ mixer and PA are considered the dominant sources of non-linearities
in the FD transceiver. Accordingly, the non-linear effect of other
transceiver components, such as the DAC, ADC, variable gain amplifier
(VGA), and LNA, is neglected. Furthermore, due to the use of a shared
local oscillator (LO) for both transmitter and receiver, the effect
of the phase noise is also ignored {[}\ref{Y. Kurzo, A. T. Kristensen, A. Burg, A. Balatsoukas-Stimming}{]}.
As such, the digital transmitted signal $x(n)$ is firstly converted
from the digital to analog form using the DAC. The analog signal is
then filtered using a low pass filter (LPF) and mixed with the carrier
signal at the IQ mixer. The IQ mixer adds a non-linear distortion
to the input signal due to the IQ imbalance, and the digital equivalent
of the IQ mixer signal can be written as {[}\ref{D.-Korpi,-L.}{]} 

\vspace{-4.5mm}
\begin{equation}
x_{IQ}(n)=\frac{1}{2}(1+\psi e^{j\theta})\,x(n)+\frac{1}{2}(1-\psi e^{j\theta})\,x^{*}(n),\label{eq:1}
\end{equation}
where $\psi$ and $\theta$ denote the transmitter's gain and phase
imbalance parameters, respectively. The mixed signal is then amplified
using the PA, which further distorts the input signal by adding additional
non-linearities. In this paper, we consider the parallel Hammerstein
model to approximate the non-linear distortion of the PA {[}\ref{D.-Korpi,-L.}{]}.
Subsequently, the output signal of the PA can be expressed as follows
{[}\ref{D.-Korpi,-L.}{]}, {[}\ref{A.-Balatsoukas-Stimming,-"Non-li}{]}: 

\vspace{-4.5mm}

\begin{equation}
x_{P\hspace{-0.5mm}A}(n)=\sum_{\underset{p\,odd}{p=1,}}^{P}\sum_{m=0}^{M_{P\hspace{-0.5mm}A}}h{}_{m,p}\thinspace x_{IQ}(n-m)^{\frac{p+1}{2}}x_{IQ}^{*}(n-m)^{\frac{p-1}{2}},\label{eq:2}
\end{equation}
where $h_{m,p}$ represents the impulse response of the parallel Hammerstein
model, while $P$ and $M_{P\hspace{-0.5mm}A}$ are the non-linearity
order and memory length of the PA, respectively.

The amplified signal is then leaked into the receiver via the SI channel
forming the SI signal. With the assumption that the FD system does
not receive any signal from any remote FD nodes (i.e., no signal of
interest is considered) and there is no thermal noise, only the SI
signal will go through the receiver. The received SI signal is firstly
filtered by the band pass filter (BPF), then amplified by the LNA,
down-converted by means of the IQ mixer, and finally converted to
digital form using the ADC. The SI signal at the receiver output is
expressed as

\vspace{-4.5mm}

\begin{equation}
y(n)=\sum_{\underset{p\,odd}{p=1,}}^{P}\sum_{q=0}^{p}\sum_{m=0}^{M-1}h{}_{m,q,p}\,x(n-m)^{q}x^{*}(n-m)^{p-q},\label{eq:3}
\end{equation}
where $h_{m,q,p}$ indicates the impulse response of a channel including
the overall effect of the PA, IQ mixer, and SI channel, while $M$
denotes the memory effect introduced to the input signal by the PA
and SI channel.

The main aim of the digital canceler is to generate an accurate estimated
version $\hat{y}(n)$ of the SI signal $y(n)$ at the receiver. Therefore,
to perform the digital SI cancellation, $\hat{y}(n)$ is subtracted
from $y(n)$, and the residual amount of the SI is approximated as
$y_{r}(n)=y(n)-\hat{y}(n)$. Hence, the amount of SI cancellation
can be given in dB as

\vspace{-3mm}
\begin{equation}
\mathit{\varPsi}_{dB}=10\log_{10}\left(\sum_{n}\left|y(n)\right|^{2}/\sum_{n}\left|y_{r}(n)\right|^{2}\right)\negmedspace.\label{eq:4}
\end{equation}

\vspace{-4mm}
\begin{figure*}
\begin{centering}
\vspace{-3.5mm}
\includegraphics[scale=0.15]{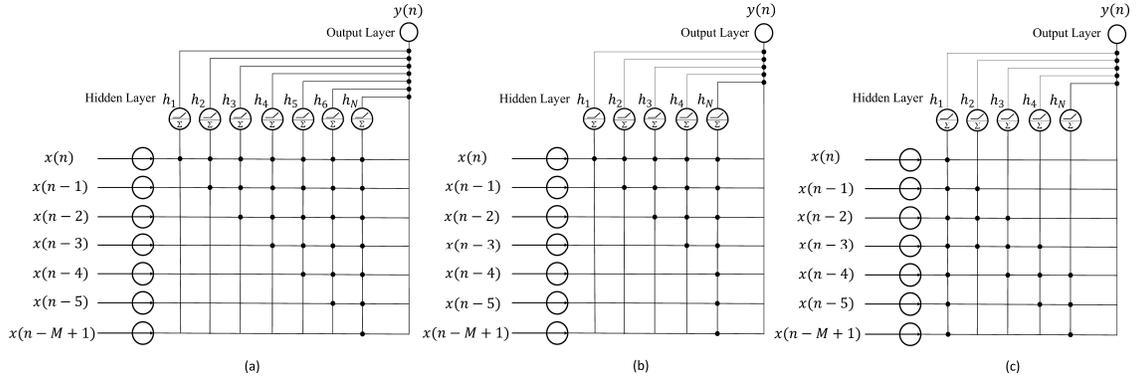}
\par\end{centering}
\centering{}\caption{\label{fig:Proposed-LWGS-neural}Proposed NNs (a) LWGS for $N=M$
(b) LWGS for $N<M$ (c) MWGS.}
\end{figure*}

\section{Proposed NN-Based Cancelers }

Cascade forward NN is an NN architecture that utilizes additional
connections from the input and every pre-layer to every post-layer
{[}\ref{B.-Warsito,-R.}{]}. The cascade forward NN has been broadly
utilized to model the time series data, and it is shown to work well
in a wide variety of problems {[}\ref{B.-Warsito,-R.}{]}. A similar
NN that employs a cascade structure is the cascade correlation NN
(CasCor NN) {[}\ref{S.-E.-Fahlman}{]}, a promising solution to speed
up the learning algorithms of the conventional NNs, such as the back-propagation.
CasCor NN starts with a simple network topology that contains only
input and output units, and then successively adds hidden units one
by one until the desired level of network error is accomplished. The
resulting network is formed in a grid topology in which each new added
unit is connected to the input and other layers' units in a cascade
structure fashion.

CasCor NN has a faster learning capability than the conventional NNs
that apply back-propagation algorithms. Moreover, it is not mandatory
in CasCor NN to determine the network structure before the training
phase since the network automatically determines its optimum configuration
{[}\ref{S.-E.-Fahlman}{]}. However, the major disadvantage of CasCor
NN is that it potentially overfits to the training data in the sense
that it yields a better performance on the training set while achieving
worse performance on a previously unseen (i.e., new) data {[}\ref{F.-S`mieja,-=00201CNeural-network}{]}.
As a result, modified versions of CasCor NN have been introduced to
avoid the overfitting of CasCor NN by applying simplified grid structures
with only partial connections in the grid {[}\ref{T.-Y.-Kwok-and}{]}.
Based on this, for the SI cancellation, various grid structures can
be investigated to model the memory effect introduced to the SI signal
in order to achieve a desired cancellation performance with reduced
computational complexity. 

Motivated by this promising idea, two novel low complexity NN structures,
named as the LWGS and MWGS, are proposed. The proposed NNs employ
a grid topology with partial connections among the different neurons
in the input and hidden layers. The main difference between the LWGS
and MWGS lies in the way utilized by each structure to pass the buffered
samples of the input signal to the different neurons in the grid in
order to efficiently simulate the SI signal\textquoteright s memory
effect. The key ideas and network structures of the proposed methods
are presented in detail in the following subsections. 

\vspace{-2mm}

\subsection{Ladder-Wise Grid Structure (LWGS) }

To imitate the memory effect introduced to the SI signal, the LWGS
is proposed as shown in Figs. \ref{fig:Proposed-LWGS-neural}(a) and
(b). The LWGS employs a grid structure similar to that used in the
CasCor NN. However, the LWGS uses the standard back-propagation technique
to minimize the network\textquoteright s error and cannot determine
its own structure as the CasCor NN. Accordingly, in the LWGS, the
network structure is selected empirically before training to achieve
the target network performance. 

The basic idea behind the LWGS is to feed the buffered data to the
network neurons in a stair-case manner as depicted in Figs. \ref{fig:Proposed-LWGS-neural}(a),
(b). Here, we denote the number of hidden units by $N$. As such,
in Fig. \ref{fig:Proposed-LWGS-neural}(a), we consider the case when
the number of hidden units is equal to the number of input units (e.g.,
$N=M=7$), where $M$ is the memory length as stated before. Starting
with the stair base, the instantaneous sample $x(n)$ is passed to
all the neurons, and every predecessor sample (i.e., $x(n-1)$, $x(n-2)$,
... etc.) is passed to a fewer number of neurons gradually. The oldest
sample, $x(n-M+1)$, which is the least one related to the current
sample $x(n)$, is allowed to be passed to only one neuron side by
side with its followers. In this manner, each neuron receives the
instantaneous sample plus part of the buffered samples to learn the
temporal behavior of the SI signal, and the outputs of all neurons
are then combined to figure out the detected pattern. Following this
approach, the LWGS could model the SI signal with only partial connections
in the grid, and therefore it can result in a significant reduction
in the computational complexity.

Furthermore, the LWGS can learn the memory effect introduced to the
SI signal with fewer connections between the input and hidden layers'
neurons. Reducing the number of connections can be done by considering
a shorter length of the ladder base (i.e., reducing the number of
neurons to be less than the memory length ($N<M$)) as shown in Fig.
\ref{fig:Proposed-LWGS-neural}(b). The idea of this configuration
is to enable the recent delayed samples that are more related to the
instantaneous sample $x(n)$ to be learned using many neurons; however,
the other samples that are less related to $x(n)$ are learned by
only one neuron (e.g., $x(n-4)$, $x(n-5),$ $x(n-M+1))$ in Fig.
\ref{fig:Proposed-LWGS-neural}(b). This will slightly degrade the
performance of the LWGS while providing a significant reduction in
the computational complexity compared to the case when $N=M.$ 

In the proposed method, the SI cancellation is performed in the digital
domain in which the non-linear part of the digital SI cancellation
signal is reconstructed using the LWGS canceler. However, the linear
part of the cancellation signal is estimated using the conventional
least square channel estimation technique where all the non-linear
effects of the different transceiver's components are neglected {[}\ref{A.-Balatsoukas-Stimming,-"Non-li}{]}.
The total cancellation achieved by the LWGS canceler is then computed
by summing the linear and non-linear cancellations {[}\ref{A.-Balatsoukas-Stimming,-"Non-li}{]},
{[}\ref{Y. Kurzo, A. T. Kristensen, A. Burg, A. Balatsoukas-Stimming}{]}. 

The effect of varying the number of hidden layer's neurons on the
cancellation performance of the LWGS is shown in Fig. \ref{Figure (3) fig:SI-cancellation-boxplots.}(a).\footnote{\label{fn:Foot note}The results in Fig. \ref{Figure (3) fig:SI-cancellation-boxplots.}
are obtained using the simulation parameters in Section \ref{sec:Results-and-Discussion}. } We test the LWGS using $N=9,10,11,12$ and depict the boxplots of
total cancellation achieved by various configurations using 20 seed
initializations. The LWGS shows flexible settings that suit different
applications. For example, moving from LWGS with nine neurons (i.e.,
LWGS (9)) to twelve neurons (i.e., LWGS (12)) augments the SI cancellation
from 44.50 to 44.75 dB; however, the increased number of neurons would
result in an increased computational complexity.

\vspace{-3mm}

\subsection{Moving-Window Grid Structure (MWGS) }

\vspace{-0.5mm}

An alternative approach that can accommodate the memory effect of
the SI signal is the moving window technique, generally recognized
as an effective method for time series prediction {[}\ref{D. Ruta, B. Gabrys, and C. Lemke}{]}.
As such, we consider the moving window with a grid topology to form
the MWGS. Similar to the LWGS, the MWGS applies the standard back-propagation
technique to minimize the network\textquoteright s error. Further,
the MWGS takes advantage of the reduced connections in the network
grid. However, in the MWGS, the considered samples of the input signals
learned by different neurons are partitioned based on a fixed-length
sliding window technique as depicted in Fig. \ref{fig:Proposed-LWGS-neural}(c).
More specifically, all the input samples are passed to the first neuron.
The main purpose of this neuron is to learn the dependencies between
all the delayed samples of the input signal. Moreover, the other employed
neurons are allowed to assist in learning the memory effect by considering
the windowed data only. Besides, sliding the window over different
neurons allows to consider all the buffered samples caused by the
non-linearities of the aforementioned FD components. For example,
in Fig. \ref{fig:Proposed-LWGS-neural}(c), a window size $W=3$ is
employed. Therefore, the first, second, and third delayed version
of $x(n)$ (i.e., $x(n-1)$, $x(n-2)$, $x(n-3))$ are considered
by the second neuron, while $x(n-2)$, $x(n-3)$, and $x(n-4)$ samples
are recognized by the third neuron, and so forth.

\begin{figure*}
\begin{centering}
\subfloat[LWGS versus different number of neurons.]{\begin{centering}
\includegraphics[scale=0.4]{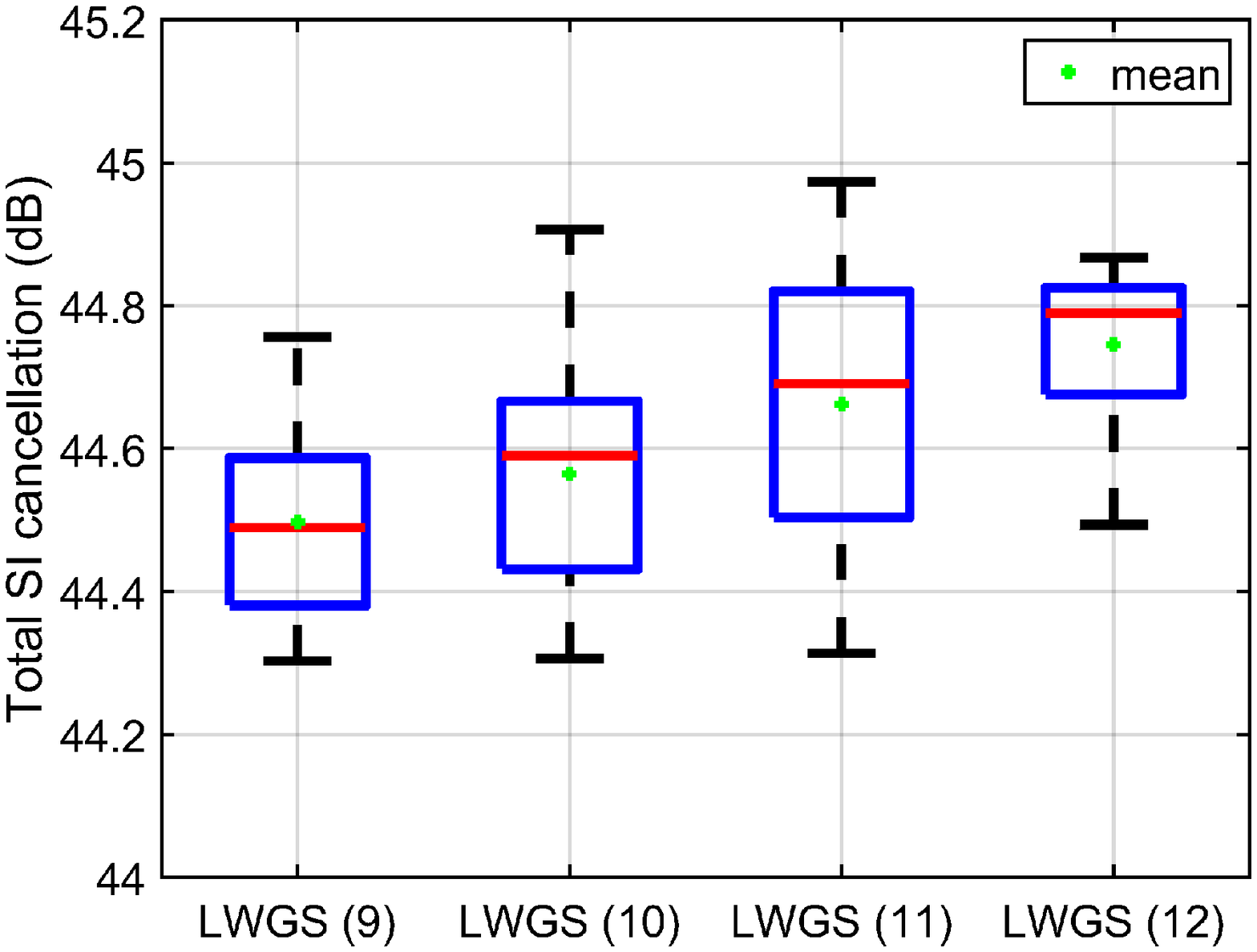}
\par\end{centering}
\centering{}} \subfloat[MWGS versus different number of neurons and window sizes.]{\begin{centering}
\includegraphics[scale=0.4]{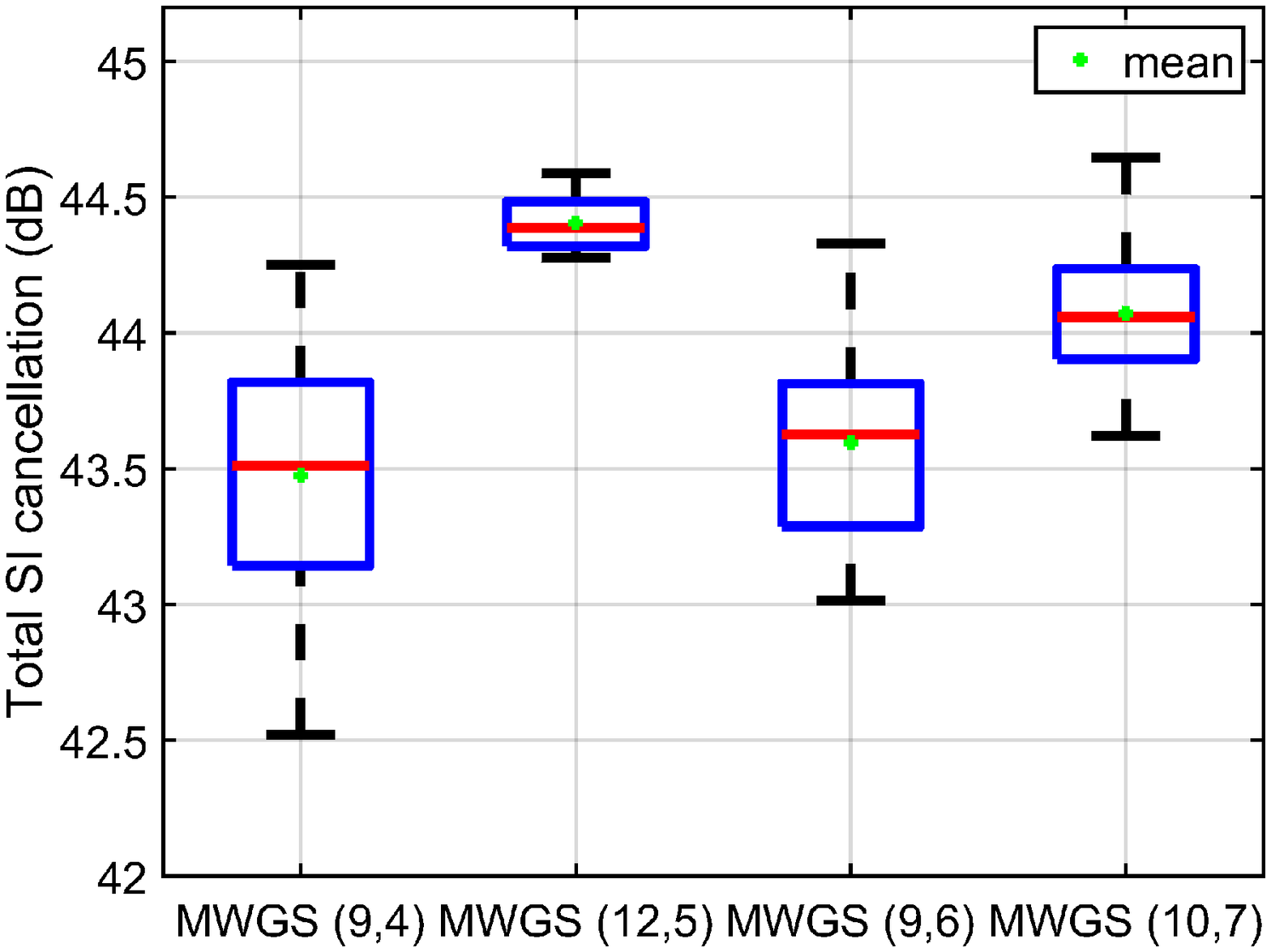}
\par\end{centering}
\centering{}}
\par\end{centering}
\caption{\label{Figure (3) fig:SI-cancellation-boxplots.}SI cancellation boxplots.}

\end{figure*}

The effect of varying the number of hidden layer's neurons $N$ and
the window size $W$ on the cancellation performance of the MWGS is
studied as well.\textsuperscript{\ref{fn:Foot note} }In this study,
the network structures of the MWGS are selected empirically based
on a trial and error approach. As such, we test the values of $N=9,10,11,12$
and $W=4,5,6,7$. Due to having many combinations between $N$ and
$W$, in Fig. \ref{Figure (3) fig:SI-cancellation-boxplots.}(b),
we only show the best four structures that achieve the highest SI
cancellation. As seen from the figure, the MWGS using twelve neurons
and window size $W=5$ (i.e., MWGS (12,5)) attains the highest SI
cancellation among the competing structures.\textcolor{blue}{{} }

\vspace{-3mm}

\section{Computational Complexity }

In this paper, we consider the total number of FLOPs as an indicator
of the computational complexity of the NN-based cancelers. The total
number of FLOPs is evaluated by calculating the total number of real-valued
operations used in the NN\textquoteright s inference process as $\varXi=\varXi_{w,b}+\varXi_{a}$
where $\varXi_{w,b}=\xi_{w,b}^{(R_{m})}+\xi_{w,b}^{(R_{a})}$ is the
sum of real-valued multiplications and additions, which account for
multiplying the input/previous layer values by the weight matrix and
adding the bias terms. Similarly, $\varXi_{a}=\xi_{a}^{(R_{m})}+\xi_{a}^{(R_{a})}$
represents the sum of real-valued multiplications and additions required
to evaluate the activation functions in the hidden layer's neurons. 

In the LWGS and MWGS NNs, the inputs, hidden layer values, and network
parameters (e.g., weights and biases) are complex-valued numbers.
Therefore, converting the complex-valued multiplications and additions
to their real-valued equivalents is required. By employing the reduced
multiplications approach {[}\ref{Y. Kurzo, A. T. Kristensen, A. Burg, A. Balatsoukas-Stimming}{]},
a complex-valued multiplication requires three real multiplications
and five real additions. Moreover, since each complex-valued addition
is implemented using two real additions, $\xi_{w,b}^{(R_{m})}$ and
$\xi_{w,b}^{(R_{a})}$ can be expressed as

\vspace{-2mm}

\begin{equation}
\xi_{w,b}^{(R_{m})}=3\xi_{w,b}^{(C_{m})},\label{eq:6}
\end{equation}

\vspace{-3mm}

\begin{equation}
\xi_{w,b}^{(R_{a})}=5\xi_{w,b}^{(C_{m})}+2\xi_{w,b}^{(C_{a})},\label{eq:7}
\end{equation}
where $\xi_{w,b}^{(C_{m})}$ and $\xi_{w,b}^{(C_{a})}$ represent
the number of complex-valued multiplications and additions, respectively,
which account for handling the weights and biases operations.

In the LWGS, $\xi_{w,b}^{(C_{m})}$ and $\xi_{w,b}^{(C_{a})}$ can
be calculated as 

\vspace{-3mm}

\begin{equation}
\xi_{w,b}^{(C_{m})}=\xi_{w,b}^{(C_{a})}=\sum_{i=1}^{N}i+M.\label{eq:8}
\end{equation}
Further, in the MWGS, $\xi_{w,b}^{(C_{m})}$ and $\xi_{w,b}^{(C_{a})}$
can be obtained as

\vspace{-2mm}

\begin{equation}
\xi_{w,b}^{(C_{m})}=\xi_{w,b}^{(C_{a})}=M+W(N-1)+N.\label{eq:9}
\end{equation}
However, $\xi_{w,b}^{(C_{m})}$ and $\xi_{w,b}^{(C_{a})}$ for CV-FFNN
can be expressed as 

\vspace{-2mm}

\begin{equation}
\xi_{w,b}^{(C_{m})}=\xi_{w,b}^{(C_{a})}=N(M+1).\label{eq:10}
\end{equation}

The proposed LWGS and MWGS employ the complex RELU ($\mathbb{C\textrm{RELU}}$)
activation function, which is defined as {[}\ref{A.-T.-Kristensen,}{]} 

\vspace{-2mm}

\begin{equation}
\varPhi(z)=\max(0,\Re(z))+j\max(0,\Im(z)),\label{eq:11}
\end{equation}
where $\Re(z)$ and $\Im(z)$ denote the real and imaginary parts
of $z$, respectively. The implementation of $\mathbb{C\textrm{RELU}}$
activation function (\ref{eq:11}) requires two real multiplications
and two complex additions (i.e., four real additions) to evaluate
the real and imaginary parts of $z$. Further, to evaluate the $\max(0,\Re(z))$
and $\max(0,\Im(z))$, two multiplexers and two comparators are required.
Herein, if we assume that each comparator comes with no cost and each
multiplexer costs one real addition {[}\ref{A.-Balatsoukas-Stimming,-"Non-li}{]},
the implementation of $\mathbb{C\textrm{RELU}}$ activation function
requires two real multiplications and six real additions.\footnote{We note that the activation functions' complexity in {[}\ref{A.-T.-Kristensen,}{]}
is evaluated by counting their usage in the hidden layer's neurons,
which is not exact.} As such, the number of real-valued multiplications $\xi_{a}^{(R_{m})}$
and additions $\xi_{a}^{(R_{a})}$ utilized for evaluating the activation
functions in the hidden layer's neurons of the LWGS and MWGS can be
given by $\xi_{a}^{(R_{m})}=2N$ and $\xi_{a}^{(R_{a})}=6N$, respectively. 

\vspace{-2mm}

\section{\label{sec:Results-and-Discussion}Results and Discussion}

In this section, we assess the performance of the LWGS and MWGS NN-based
cancelers in terms of mean square error (MSE), SI cancellation, and
computational complexity. In addition, a comparison between the proposed
methods and the NN-based cancelers in the literature is also investigated.
All the considered NNs are trained using complex-valued inputs and
implemented in Python using Keras library and TensorFlow back-end.
In this work, we examine the use of the measured dataset presented
in {[}\ref{A.-Balatsoukas-Stimming,-"Non-li}{]} and {[}\ref{A.-T.-Kristensen,}{]}.
Hence, we train and test the NNs using measured data from a realistic
FD testbed, which applies an orthogonal frequency division multiplexing
signal with 1024 sub-carriers using a quadrature phase-shift keying
modulation and 10 MHz pass-band bandwidth. The dataset, containing
20,480 samples, is split into a training set that consists of 90\%
of samples and a testing set that includes the remaining 10\%. We
adopt the back-propagation technique, Adam optimization, and $\mathbb{C\textrm{RELU}}$
activation function for the NNs {[}\ref{A.-T.-Kristensen,}{]}. The
networks\textquoteright{} hyperparameters, such as the batch size
and learning rate, are tuned using five-fold cross-validation to select
their optimal values. Based on cross-validation, we employ a learning
rate of 0.0045 and a batch size of 62 to train the NNs. Besides, we
consider $M=13$ for the polynomial and NN-based cancelers {[}\ref{A.-Balatsoukas-Stimming,-"Non-li}{]},
{[}\ref{A.-T.-Kristensen,}{]}. 

All the NN-based cancelers are employed to model the non-linear part
of the SI signal. Furthermore, for the sake of comparison, the NNs
settings are selected in such a way that they achieve a similar cancellation
performance to the polynomial canceler with $P=5$. The polynomial
canceler at $P=5$ produces 44.45 dB cancellation and requires 1556
FLOPs and 312 network parameters to be implemented {[}\ref{A.-T.-Kristensen,}{]}.
As such, to achieve the target cancellation of the polynomial canceler,
the CV-FFNN requires at least a single hidden layer with seven neurons
(i.e., CV-FFNN (7)) {[}\ref{A.-T.-Kristensen,}{]}. In addition, from
Fig. \ref{Figure (3) fig:SI-cancellation-boxplots.}, it is observed
that the LWGS (9) and LWGS (10) achieve the target cancellation as
they provide 44.50 and 44.56 dB, respectively. Further, the MWGS (12,5)
attains 44.40 dB, which is very close to the target cancellation.
Thus, in this analysis, we consider CV-FFNN (7), LWGS (9), LWGS (10),
and MWGS (12,5) as promising NN-based cancelers that can be used as
alternatives to the traditional polynomial canceler. 

\begin{figure}
\begin{centering}
\subfloat[MSE performance.]{\begin{centering}
\includegraphics[scale=0.407]{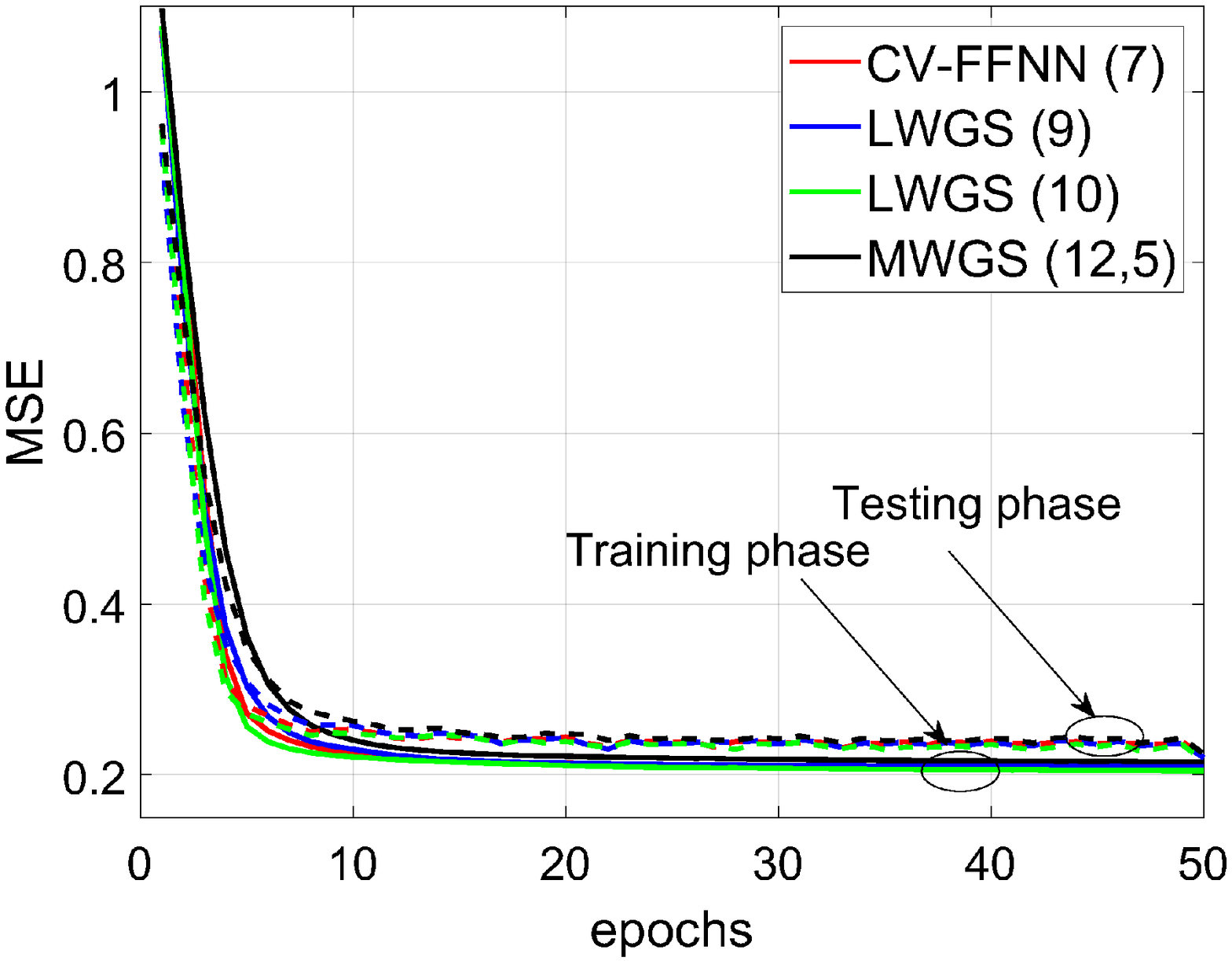}
\par\end{centering}
\centering{}}\subfloat[SI cancellation boxplots.]{\begin{centering}
\includegraphics[scale=0.4]{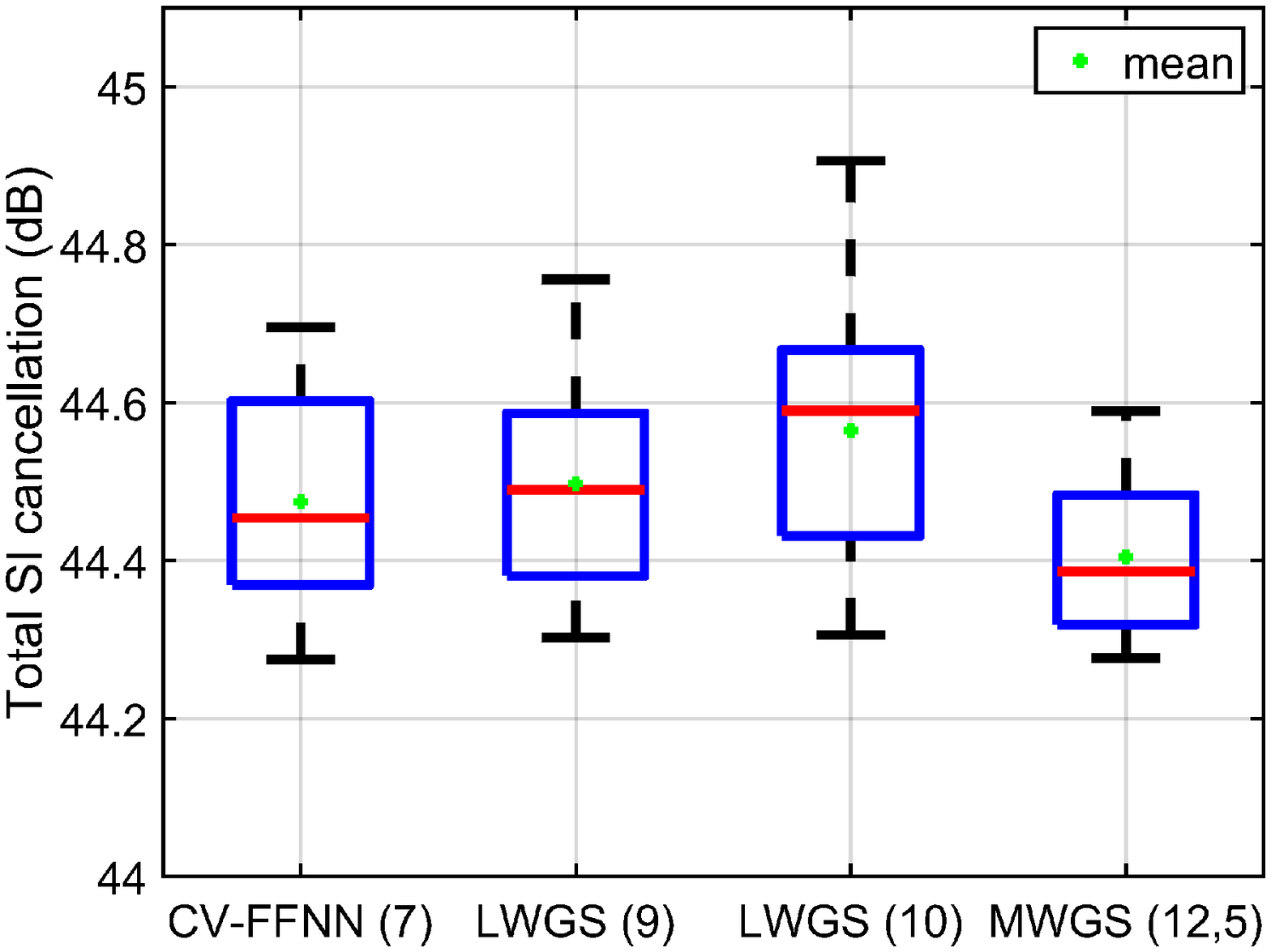}
\par\end{centering}
\centering{}}
\par\end{centering}
\caption{\label{Figure (4)fig:Performance-comparison-for}Performance comparison
for different network structures. }
\end{figure}

In Fig. \ref{Figure (4)fig:Performance-comparison-for}(a), the MSE
values of the aforementioned NNs are evaluated on the training and
testing data, respectively, using 20 seed initializations. As seen
from the figure, the considered NNs achieve a comparable MSE for the
target cancellation performance. Fig. \ref{Figure (4)fig:Performance-comparison-for}(b)
depicts the boxplots of SI cancellation achieved by the considered
NN-based cancelers using the above-selected settings. It is apparent
from the figure that CV-FFNN (7), LWGS (9), and MWGS (12,5) attain
a comparable cancellation performance to the polynomial canceler.
However, LWGS (10) provides a slightly higher SI cancellation. It
is worth noting that the LWGS structure slightly outperforms the cancellation
of the MWGS as it passes the instantaneous sample $x(n)$ (i.e., most
significant sample) to all neurons, which enables it to learn the
SI signal's temporal behavior better than the MWGS.

The complexity analysis for the different NN-based cancelers is provided
in Table. \ref{tab:Complexity-reduction-for}, where the polynomial
canceler complexity is considered as reference, and the NNs complexity
is computed in terms of the number of FLOPs and the number of network
parameters used to perform the total SI cancellation (i.e., linear
and non-linear cancellations). As seen from the table, the LWGS (9)
reduces the number of FLOPs by more than 49\% while achieving a similar
cancellation to the polynomial-based canceler. Furthermore, the LWGS
(10) outperforms the cancellation performance of the polynomial canceler
while requiring 7\% more FLOPs than the LWGS (9). Accordingly, the
proposed LWGS provides a flexible trade-off between the cancellation
performance and the computational complexity. In addition, the MWGS
(12,5) saves 34\% computations compared to the polynomial-based canceler,
while the conventional CV-FFNN (7) saves only 25\% of the computations.
The previous results reveal the superiority of the proposed NNs compared
to the polynomial and state-of-the-art NN-based cancelers. 

\vspace{-3mm}

\section{Conclusion}

In this paper, two novel low complexity NN structures, namely the
ladder-wise grid structure (LWGS) and moving-window grid structure
(MWGS), are proposed to model the SI signal with low computational
complexity. The former employs a stair-based structure to accommodate
the memory effect of the SI signal. The latter uses a fixed-window
procedure to model the temporal behavior of the SI signal. Our findings
showed that the proposed LWGS and MWGS provide the same cancellation
performance of the polynomial-based canceler while attaining 49.87\%
and 34.19\% reduction in the computational complexity, respectively.
In addition, the proposed LWGS and MWGS offer superior performance
over the state-of-the-art NN-based cancelers by exhibiting 24.7\%
and 9\% complexity reduction, respectively.

\begin{table}
\vspace{5mm}
{\scriptsize{}\caption{\label{tab:Complexity-reduction-for}Complexity reduction for different
network structures compared to the polynomial model with $P=5$.}
}{\scriptsize\par}
\centering{}{\scriptsize{}}%
\begin{tabular}{cccccc}
\toprule 
\addlinespace
\multirow{2}{*}{{\scriptsize{}Network}} & \multirow{2}{*}{{\scriptsize{}Cancellation}} & \multicolumn{2}{c}{{\scriptsize{}Complexity}} & \multicolumn{2}{c}{{\scriptsize{}Complexity Reduction}}\tabularnewline
\cmidrule{3-6} \cmidrule{4-6} \cmidrule{5-6} \cmidrule{6-6} 
\addlinespace
 &  & {\tiny{}\# Parameters} & {\tiny{}\# FLOPs} & {\tiny{}\# Parameters} & {\tiny{}\# FLOPs}\tabularnewline
\midrule
{\scriptsize{}Polynomial ($P\negthinspace=$\,5)} & {\scriptsize{}44.45 dB} & {\scriptsize{}312} & {\scriptsize{}1556} & {\scriptsize{}-} & {\scriptsize{}-}\tabularnewline
{\scriptsize{}CV-FFNN (7)} & {\scriptsize{}44.47 dB} & {\scriptsize{}238} & {\scriptsize{}1164} & {\scriptsize{}-23.72\%} & {\scriptsize{}-25.19\%}\tabularnewline
{\scriptsize{}LWGS (9)} & {\scriptsize{}44.50 dB} & {\scriptsize{}162} & {\scriptsize{}780} & {\scriptsize{}-48.08\%} & {\scriptsize{}-49.87\%}\tabularnewline
{\scriptsize{}LWGS (10)} & {\scriptsize{}44.56 dB} & {\scriptsize{}184} & {\scriptsize{}888} & {\scriptsize{}-41.03\%} & {\scriptsize{}-42.93\%}\tabularnewline
{\scriptsize{}MWGS (12,5)} & {\scriptsize{}44.40 dB} & {\scriptsize{}212} & {\scriptsize{}1024} & {\scriptsize{}-32.05\%} & {\scriptsize{}-34.19\%}\tabularnewline
\bottomrule
\end{tabular}\vspace{-5mm}
\end{table}

\end{document}